\documentstyle[psfig]{mn}

\newif\ifAMStwofonts

\ifoldfss
  \ifCUPmtlplainloaded \else
    \NewTextAlphabet{textbfit} {cmbxti10} {}
    \NewTextAlphabet{textbfss} {cmssbx10} {}
    \NewMathAlphabet{mathbfit} {cmbxti10} {} 
    \NewMathAlphabet{mathbfss} {cmssbx10} {} 
  \fi
  \ifAMStwofonts
    \ifCUPmtlplainloaded \else
      \NewSymbolFont{upmath} {eurm10}
      \NewSymbolFont{AMSa} {msam10}
      \NewMathSymbol{\upi}     {0}{upmath}{19}
      \NewMathSymbol{\umu}     {0}{upmath}{16}
      \NewMathSymbol{\upartial}{0}{upmath}{40}
      \NewMathSymbol{\leqslant}{3}{AMSa}{36}
      \NewMathSymbol{\geqslant}{3}{AMSa}{3E}

    \fi
  \fi
\fi 

\ifnfssone
  \newmathalphabet{\mathit}
  \addtoversion{normal}{\mathit}{cmr}{m}{it}
  \addtoversion{bold}{\mathit}{cmr}{bx}{it}
  \newmathalphabet{\mathbfit} 
  \addtoversion{normal}{\mathbfit}{cmr}{bx}{it}
  \addtoversion{bold}{\mathbfit}{cmr}{bx}{it}
  \newmathalphabet{\mathbfss} 
  \addtoversion{normal}{\mathbfss}{cmss}{bx}{n}
  \addtoversion{bold}{\mathbfss}{cmss}{bx}{n}
  \ifAMStwofonts
    \ifCUPmtlplainloaded \else
      \UseAMStwoboldmath
      \makeatletter
      \new@mathgroup\upmath@group
      \define@mathgroup\mv@normal\upmath@group{eur}{m}{n}
      \define@mathgroup\mv@bold\upmath@group{eur}{b}{n}
      \edef\UPM{\hexnumber\upmath@group}
      \new@mathgroup\amsa@group
      \define@mathgroup\mv@normal\amsa@group{msa}{m}{n}
      \define@mathgroup\mv@bold\amsa@group{msa}{m}{n}
      \edef\AMSa{\hexnumber\amsa@group}
      \makeatother
      \mathchardef\upi="0\UPM19
      \mathchardef\umu="0\UPM16
      \mathchardef\upartial="0\UPM40
      \mathchardef\leqslant="3\AMSa36
      \mathchardef\geqslant="3\AMSa3E
    \fi
  \fi
\fi 

\ifnfsstwo
  \DeclareMathAlphabet{\mathbfit}{OT1}{cmr}{bx}{it}
  \SetMathAlphabet\mathbfit{bold}{OT1}{cmr}{bx}{it}
  \DeclareMathAlphabet{\mathbfss}{OT1}{cmss}{bx}{n}
  \SetMathAlphabet\mathbfss{bold}{OT1}{cmss}{bx}{n}
  \ifAMStwofonts
    \ifCUPmtlplainloaded \else
      \DeclareSymbolFont{UPM}{U}{eur}{m}{n}
      \SetSymbolFont{UPM}{bold}{U}{eur}{b}{n}
      \DeclareSymbolFont{AMSa}{U}{msa}{m}{n}
      \DeclareMathSymbol{\upi}{0}{UPM}{"19}
      \DeclareMathSymbol{\umu}{0}{UPM}{"16}
      \DeclareMathSymbol{\upartial}{0}{UPM}{"40}
      \DeclareMathSymbol{\leqslant}{3}{AMSa}{"36}
     \DeclareMathSymbol{\geqslant}{3}{AMSa}{"3E}
    \fi
  \fi
\fi 

\ifCUPmtlplainloaded \else
  \ifAMStwofonts \else 
    \def\upi{\pi}
    \def\umu{\mu}
    \def\upartial{\partial}
  \fi
\fi

\title{The nature of the Wolf-Rayet galaxy Mrk 209 from photoionization models}
\author[E. P{\'e}rez-Montero \& A.I. D{\'\i}az]
       {Enrique-P{\'e}rez-Montero$^{1,2}$\thanks{Post-Doc fellow of Ministerio de Educaci\'on y
    Ciencia, Spain; enrique.perez@ast.obs-mip.fr} \& \'Angeles I. D\'{\i}az$^{2}$\thanks{On sabbatical leave at IoA, Cambridge}\\ 
$^{1}$ Institute d'Astrophysique de Toulouse-Tarbes. Observatoire Midi-Pyr\'en\'ees. 14, avenue Edouard Belin. 31400. Toulouse. France\\ 
 $^{2}$ Departamento de F\'{\i}sica Te\'orica, C-XI, Universidad Aut\'onoma de
Madrid, 28049 Madrid, Spain\\  }
        
\date{Accepted 
      Received ;
      in original form November 2006}

\pagerange{\pageref{firstpage}--\pageref{lastpage}}
\pubyear{2006}

\begin{document}

\maketitle

\label{firstpage}

\begin{abstract}
We present a detailed photoionization model of the brightest knot of star formation in the blue compact dwarf galaxy Mrk 209. The model reproduces the intensities of main lines emitted by the ionized gas, resulting in a very good agreement between observed and predicted  line temperatures and chemical abundances of the observed ionic species.

The model has been calculated using the spectral energy distribution of a massive cluster of recent formation as the ionizing source. The features of Wolf-Rayet stars observed in the spectrum of the object, along with its ionizing properties, lead to different solutions for the ages and characteristics of the ionizing stellar populations.The found solutions are  compatible with either a composite population of two ionizing clusters or a continuous star formation. In the first case, a young cluster, with an age less than 3 Myr, would be responsible for most of the ionization properties while an older cluster, with either 3.2 or 4.2 Myr, would be responsible for the emission of the observed WR features in the spectrum of the knot. In the second case, the duration of the star formation episode is found to be 3.6 Myr.  

The addition of a certain amout of dust was required in order to reproduce correctly the measured electron temperatures. This dust implies a gas/dust ratio of 6.13$\times$10$^{-3}$, identical to that in the Milky Way, and a fraction of absorbed photons of $f$ = 0.58. 
The taking into account the dust grain physics combined with a thick-shell geometry solves the problem of the heating in this object and allows the fitting of the auroral lines, the line temperatures and the ionization structure of the nebula with a
negligible presence of temperature fluctuations, in agreement with the most recent results found for this type of objects
from the measurement of the Balmer jump temperature. 

An analysis of the ionization structure of the gas as well as the calculation of the main ionization correction factors
for the unseen ionic stages is presented.

\end{abstract}


\section{Introduction}

Blue compact dwarf galaxies (BCD) are objects whose spectra, as in the case of H{\sc II} regions, are dominated by emission lines from the gas ionized by very recent episodes of star formation  (Sargent \& Searle, 1970).  Many intermediate redshift objects have properties similar to the BCD in the Local Universe and it is possible that they were common in the past and they have evolved to other kind of objects. Therefore, the study of the star forming history in BCDs is of great importance.

In general, BCDs are characterized by their compact aspect, very low metallicities, gas richness and blue colors 
(Kunth \& \"Ostlin, 2000). These properties have made of them
good candidates to host their first episodes of star formation. Nevertheless, recently, the detection in most of them of  low surface-brightness elliptical haloes, or the presence of stars belonging to older populations , have caused this interpretation to be reconsidered.  Nowadays, only few candidates remain controversial

There are several works in the literature whose main aim is to study the weight of these older stellar
populations in BCDs and, hence, to find out the actual age and evolutionary status of these objects.
Among these studies, in the sample of local objects, observations with enough spatial resolution to provide photometry of the individual stars have allowed, by means of colour-magnitude diagrams, to date some of the bursts ({\em e.g.} {\sc VII}Zw403 by Schulte-Ladbeck et al., 1998; {\sc I}Zw18 by Aloisi et al., 2000). In all these works evidences for stellar populations older than 500 Myr have been found. Besides, old stellar low-surface brightness components have been detected in  these local objects by studying the radial light distribution in the optical ({\em e.g.} Gil de Paz et al., 2003) and the near infrared ({\em e.g.} Noeske et al., 2005).

For more distant galaxies, the study of the stellar population is limited to the interpretation
of the integrated spectrophotometric properties by means of empirical and evolutionary population synthesis techniques. The results from this kind of works ({\em e.g.} Raimann et al., 2000) show that most 
blue compact dwarves are age-composite stellar systems, with discontinuous star-forming histories. 

Nevertheless, there is still a lack of spatially resolved studies of the properties of the ionizing stellar populations and
the ionized surrounding gas in the different knots of star formation. This work is usually approached by means of photoionization models, trying to reproduce the measured fluxes and equivalent widths of the  lines 
emitted by the ionized gas, under different assumptions about its geometry, metal content and  ionizing incident radiation. Several improvements have been produced that increase our confidence in the results from this kind of models: inclusion of blanketing effects in O and WR expanding stellar atmosphere models (Pauldrach et al., 2001; Smith et al., 2002) and recent computation of NLTE blanketed plane-paralel models (Lanz \& Hubeny, 2003). Yet, model computed evolutionary sequences show important disagreements with observations. Among others,
the [O{\sc iii}]/H$\beta$ vs [O{\sc ii}]/H$\beta$ and [O{\sc iii}]/H$\beta$ vs [O{\sc I}]/H$\beta$ relations 
are not well reproduced by evolutionary model sequences in the sense that predicted
collisionally excited lines result too weak compared to observations, which would require 
additional heating mechanisms or the presence of chemical inhomogeneities (Stasi\'nska \& Izotov, 2003) .

These facts seem to point out to a still incomplete description of the ionization structure of the emitting nebulae which would improve with the determination of more than one line electron temperature. This has been attempted recently by several authors ({\em e.g.} Garnett, 1989; P\'erez-Montero \& D\'\i az, 2003, hereinafter PMD03; H\"agele et al. 2006) by enlarging the spectral sampling to include the lines of [O{\sc ii}]  at $\lambda\lambda$ 7319,30 \AA\ and [S{\sc iii}]  at $\lambda\lambda$ 9069, 9532 \AA . The far-red [S{\sc iii}] emission lines can provide better estimates of the  ionization parameter of the gas, via the [S{\sc iii}]/[S{\sc ii}] quotient (D\'{\i}az et al., 1991), and the equivalent effective temperature of the ionizing clusters, via the $\eta$ parameter (V\'{\i}lchez \& Pagel, 1988), than the strong widely used optical lines of oxygen, mainly due to their lower dependence on other quantities, such as age or metallicity.
These parameters are key factors to provide adequate input constraints for photoionization models. 

Although the detection and analysis of the near-infrared emission lines is hampered by the strong and variable absorption features of the Earth's atmosphere, a thorough atmospheric extinction correction and a good spectral resolution aid to remove the sky background providing a good level of confidence in the measurement of the far red [S{\sc iii}] lines. The number of H{\sc ii} galaxies observed both in the optical and the far-red spectral range, up to 1$\mu$, is still scarce but it is steadily increasing.

In this work, we apply a detailed photoionization model to the brightest knot of the BCD Mrk209, for which both wide range spectroscopic observations and photometric data exist. 

In the next section we describe the properties of Mrk209 as given in the literature, focusing on
the main burst of star formation, measuring their photometric properties and the gas physical conditions. In section
3 we describe the photoionization model we have used with the inclusion of the input stellar ionizing population and the effects  of dust. Section 4 presents and discusses the model results in comparison with observations.  Finally, the conclusions of this work are presented in section 5.

\section{Properties of the object}

Mrk 209, also known as IZw36 and UGCA 281,  is a blue compact dwarf galaxy that has been extensively studied. The work by Viallefond \& Thuan (1983) revealed a core-halo structure of the diffuse neutral gas, with the star-forming regions located near the
core, but slightly shifted with respect to the peak in H{\sc i} column density. They measured its 
virial mass which is about 6 times that in H{\sc i}, which could imply a large contribution of dark matter. Loose \& Thuan (1986) discovered an extended, elliptical background sheet underlying the compact, actively star forming core and they classified this galaxy as {\em iE}, corresponding to blue compact dwarves with an irregular burst of star formation in its core, surrounded by regular elliptical isophotes.

Although Fanelli et al. (1988), from an IUE UV spectral study, pointed to Mrk209 being a young galaxy
undergoing its first episode of star formation, later works have ruled out this conjecture. The red colours observed by Deharveng (1994) and by Papaderos et al. (1996) require the presence of old stellar populations in  the elliptical host galaxy. Schulte-Ladbeck et al. (2001) were able, by means of HST observations with 
FOC in the optical and NICMOS in the near-IR, to build a colour-magnitude diagram identifying individual stars of the red giant branch (RGB), some of them with ages between 1 and 2 Gyr. These authors claim for an almost-continuous star formation history for this galaxy with inactive periods never longer than 100 Myr.

Gil de Paz et al. (2003; hereinafter GMP03) have produced photometric observations in H$\alpha$ and the B and R bands. They find that the H$\alpha$ emission peaks at a region towards the west of the galaxy,  accompanied by two weaker star forming regions towards the east. The weakest of these regions 
lacks near infrared emission (Noeske et al., 2005) implying that the dominant population is still very young.

Regarding the brightest knot of star formation, Deharveng et al. (1994) showed that their young stars
have an age less than 12 Myr. Later evolutionary synthesis models by Mas-Hesse \& Kunth (1999) have lowered this upper limit further to only $\approx$ 2.7 Myr. Besides, Izotov et al. (1997; hereinafter ITL97) and Schaerer et al. (1999) discovered WR features in this knot.

Spectrophotometric observations in the range between 3000 and 7800 {\AA} from ITL97 and between 7000 and 9800 {\AA} (PMD03) exist for this burst. The slit width used in both works is identical and the agreement of the relative fluxes in the coincident spectral range is good; therefore, we can be confident about the spatial coincidence of both sets of observations. The intensities and equivalent widths of the observed WR features were measured by Guseva et al. (2000; hereinafter G00). 

\begin{table}
\begin{minipage}{85mm}
\vspace{-0.3cm}
\normalsize
\caption{Properties of the main burst of star formation in Mrk 209.}

\begin{center}
\begin{tabular}{cc}
\hline
\hline
Distance (Mpc)    &    6.6$\pm$1.1   \\
F(H$\alpha$)\footnote {in units of $10^{-13}$ erg $\cdot$ s$^{-1} \cdot$cm$^{-2}$}   &   14.97$\pm$0.01 \\
Z \footnote{in mass, scaled to the oxygen solar abundance from Allende-Prieto, 2001}  &  0.0036  \\ 
C(H$\beta$)   & 0.06 \\
B (mag)\footnote{not corrected for reddening}    &   16.58$\pm$0.13 \\
B - R (mag)  &   0.28$\pm$0.18  \\
EW(H$\beta$) (\AA)  &  206 \\
EW(4650 \AA) (\AA)  &  5.01$\pm$0.10 \\
log L(4650 \AA) \footnote{in units of erg $\cdot$ s$^{-1}$}  &  37.35$\pm$0.13 \\
EW(5808 \AA) (\AA)  &  1.74$\pm$0.15 \\
log L(5808 \AA)$^e$  &  36.62$\pm$0.16 \\

\hline
\hline
\end{tabular}

\end{center}

\label{data}
\end{minipage}
\end{table}

We have collected the main photometric and spectroscopic properties of the brightest knot of star formation in Table \ref{data}. Regarding the spectroscopic data, WR feature intensities and equivalent widths (G00) refer to a slit width of 2 arcsec, while the  emission line fluxes, H$\beta$ equivalent width and reddening constant c(H$\beta$) (ITL97; PMD03) refer to a slit width of 1.5 arcsec. Besides, we have compiled ISO observations of this object (Nollenberg et al., 2002) that account for line intensities of 4.05 $\mu$ Br$\alpha$, 10.5 $\mu$ [S{\sc iv}] and 18.7 $\mu$ [S{\sc iii}].
in the mid infrared. Nevertheless the apertures used for these observations are much larger that those used in the
optical and the near infrared and, althought it is not expected to affect the fluxes of the lines of the ions of high excitation it could do it to the hydrogen recombination line. 

Luminosities have been derived assuming a distance of 6.6$\pm$1.1 Mpc (Schulte-Ladbeck et al. 2001). The quoted errors for luminosities include the uncertainty in the distance to the galaxy. Regarding the photometric data, we have selected the area of the H$\alpha$ image from GMP03 containing the same flux as in G00 corrected for aperture, and we have re-measured the B and R magnitudes in the corresponding images. The selected region has a mean radius of 72$\pm$18 pc.  The reddening corrected emission line intensities, relative to H$\beta$ = 100, are given in Table \ref{lines}.

Using these emission line data we have recalculated the electron temperatures and density, and
the ionic chemical abundances. We have used the same algorithm described in PMD03: the five-level statistical equilibrium model in the TEMDEN and IONIC tasks of the NEBULAR package in STSDAS, based on the FIVEL program developed by De Robertis et al. (1987) and improved by Shaw \& Dufour (1995). We have used the transition probabilities and collision strengths included in the photoionization code CLOUDY (version 96.0; Ferland et al., 1998), except in the case of O$^+$ for which we have used the transition probabilities from Zeippen (1982) and the collision strengths from Pradham (1976), which offer more reliable results for nebular diagnostics (Wang et al., 2004) for this species.

\section{Model description}

The model here described simulates the properties of the ionized gas and the stellar ionizing
population of the brightest knot of star formation in Mrk 209. It has been calculated using the photoionization code CLOUDY (Version 96.0, Ferland et al., 1998). This model is characterized by a set of input 
parameters including the ionizing continuum, the nebular geometry, the gas density and chemical abundances.
The main properties of the best fitted model are summarized in Table \ref{model}.

\begin{table}
\begin{minipage}{85mm}
\vspace{-0.3cm}
\normalsize
\caption{Emission line intensities, relative to H$\beta$= 100,  with their 
corresponding errors, observed and the predicted by our model and the model in PMD03.}

\begin{center}
\begin{tabular}{cccc}
\hline
\hline
 Emission line   &   Observed & This model & PMD03  \\
\hline
3727 {\AA} [O{\sc ii}]\footnote{from Izotov et al., 1997}   &  71.9$\pm$0.2  &  69.4  &  68.7 \\
3868  {\AA} [Ne{\sc iii}]$^a$  &  45.7$\pm$0.1  &  46.0 &  -- \\
4072 {\AA} [S{\sc ii}]$^a$   &  1.2$\pm$0.1  &  1.2  & -- \\
4363 {\AA} [O{\sc iii}]$^a$  &  12.7$\pm$0.1 &  12.4  & 9.6\\
4658 {\AA} [Fe{\sc iii}]$^a$ &  0.3$\pm$0.1  & 0.4 & -- \\
4959 {\AA} [O{\sc iii}]$^a$  &  196.0$\pm$0.3 &  190.1  & 200.8 \\
5007 {\AA} [O{\sc iii}]$^a$  &  554.3$\pm$0.8 &  574.2 & 604.4  \\
6584 {\AA} [N{\sc ii}]\footnote{from P\'erez-Montero \& D\'{\i}az, 2003}   &  3.6$\pm$0.1  &  3.6 & --\\
6725 {\AA} [S{\sc ii}]$^b$   &  10.6$\pm$0.4  &  10.2  & 6.8 \\ 
6312 {\AA} [S{\sc iii}]$^b$   &  1.9$\pm$0.1  &  1.9  & 1.1 \\
7137 {\AA} [Ar{\sc iii}]$^b$  &  5.5$\pm$0.1  &  5.6 & -- \\
7325 {\AA} [O{\sc ii}]$^b$   &  1.7$\pm$0.2  &  2.6  & 2.4 \\
9069 {\AA} [S{\sc iii}]$^b$   &  12.2$\pm$1.2  &  12.1  & 8.7 \\
10.5 $\mu$ [S{\sc iv}]\footnote{from Nollenberg et al., 2002} &   42.6$\pm$7.6    &   16.5     &  -- \\
18.7 $\mu$ [S{\sc iii}]$^c$ &   17.4$\pm$4.5    &   9.7      &  -- \\

\hline
\hline
\end{tabular}

\end{center}

\label{lines}
\end{minipage}
\end{table}

\begin{table}
\begin{minipage}{85mm}
\vspace{-0.3cm}
\normalsize
\caption{Properties of the modelled H{\sc ii} region.}

\begin{center}
\begin{tabular}{ccc}
\hline
\hline
                        &   Observed & This model   \\
\hline
EW(H$\beta$)  &  206 {\AA}  &  210 {\AA} \\
log Q(H)\footnote {in photons $\cdot$ s$^{-1}$}    & 51.75$\pm$0.13 & 51.96\footnote{not absorbed by dust}    \\
log $U$   &  -2.3 $\pm$ 0.1 & -2.22 \\
filling factor  &     &  0.077  \\
Inner radius  &     & 30 pc \\
Str\"omgren radius  & 72$\pm$18 pc & 48 pc \\
Absorption factor $f$ &    & 0.58 \\
Dust-to-gas ratio  &      &  6.13 $\cdot$ 10$^{-3}$ \\
C(H$\beta$) & 0.06 & 0.19 \\

\hline
\hline
\end{tabular}
\end{center}

\label{model}
\end{minipage}
\end{table}

\subsection{The ionizing stellar population}

The WR phase in the evolution of massive
stars is a short and transient stage, and therefore the presence of WR features in the spectrum of a star forming region allows a dating of the ionizing stellar population. 
The emission features usually observed consist of two broad bumps. The first, usually called as {\em blue} is
at 4650 {\AA} and is a blend of the N{\sc v} $\lambda$ 4605, 4620 {\AA}, N{\sc iii} $\lambda$ 4634, 4640 {\AA},
C{\sc iii}/C{\sc iv} $\lambda$ 4650, 4658 {\AA} and He{\sc ii} $\lambda$ 4686 {\AA} broad WR lines. These are
emitted mainly by WN and early WC stars. Besides, early WN stars emit broad emission features of C{\sc iv} at
$\lambda$ 5808 {\AA}, known as the {\em red} bump. 
Although the number of WR stars is expected to be much
lower in relation to the number of O stars at low metallicites, they have been detected 
in the spectra of some blue compact dwarfs, including Mrk209 (ITL97, Schaerer et al., 1997) implying the presence of stars
in the burst with an age between 1 and 5 Myr, that is the age of the WR stars.

\begin{figure*}
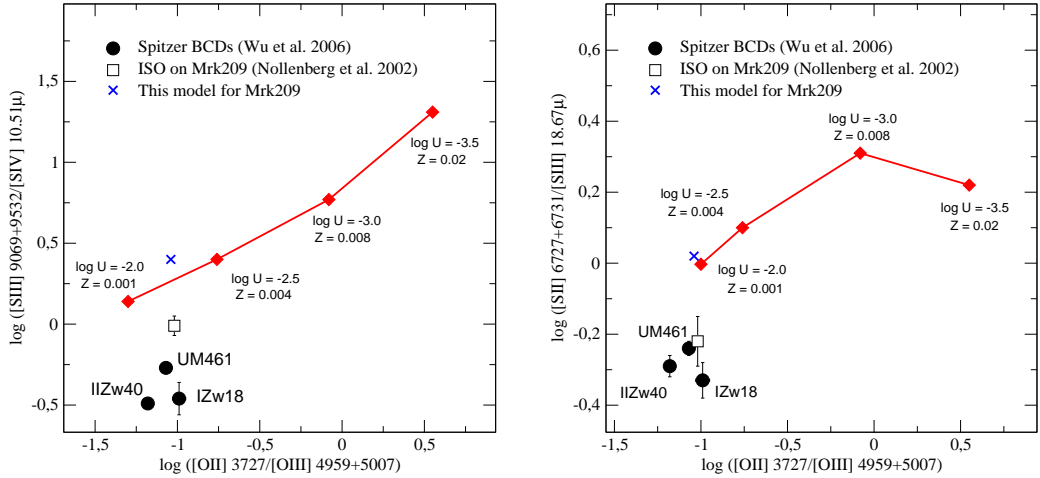

\begin{minipage}{175mm}
\centerline{
\psfig{figure=s4s3_o3o2.eps,width=6.5cm,clip=}
\hspace{0.5cm}
\psfig{figure=s3s2_o3o2.eps,width=6.5cm,clip=}}
\caption{Diagnostic diagrams of the quotients of [S{\sc iv}] 10.51 $\mu$ / [S{\sc iii}] (9069+9532 \AA), at left and of
[S{\sc iii}] 18.67 $\mu$ / [S{\sc ii}] (6717+6731 \AA), at right, versus the quotient of [O{\sc ii}] 3727 \AA / [O{\sc iii}] (4959+5007 \AA), for
a sample of Blue Compact Dwarf Galaxies observed both in the near and the mid infrared and for a sequence of
models covering the physical condicitons of ionized gaseous nebulae. All model sequences, including the model
presented here, fail to predict correctly the intensities of the sulphur emission lines in the mid-IR.}

\label{midir}
\end{minipage}
\end{figure*}

\begin{figure*}
\begin{minipage}{175mm}
\centerline{
\psfig{figure=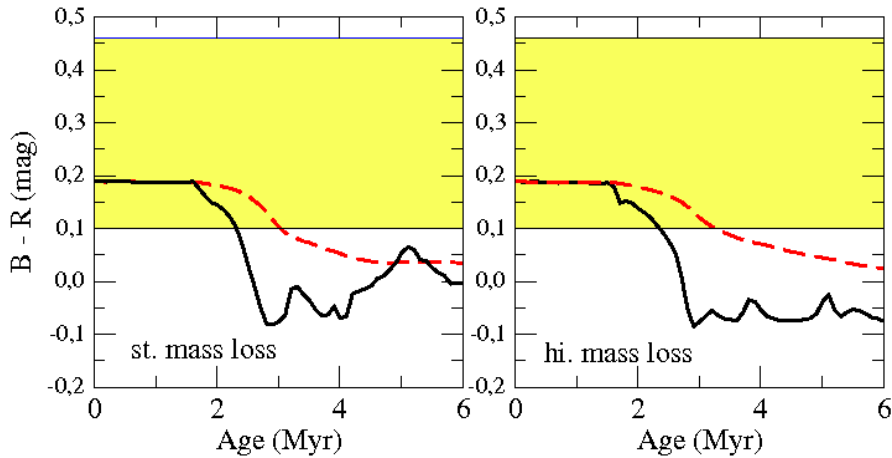,height=6.5cm,clip=}}
\caption{The colour index B-R predicted by evolutionary synthesis models for both standard and high mass loss stellar evolution and for instantaneous (thick solid line) and continuous (dashed line) star formation rates. The bands represent the values measured in the brightest star forming region in the photometric images.}
\label{br0004}
\end{minipage}
\end{figure*}

We have calculated the spectral energy distributions of ionizing star clusters in this range of ages using evolutionary synthesis techniques (Starburst 99; Leitherer et al., 1999) with the inclusion of the  implementation by Smith et al. (2002) using the model atmospheres from Pauldrach et al. (2001) and Hillier \& Miller (1998) for O and WR stars respectively which include the effects of blanketing. The used initial mass function is a Salpeter's one, with masses between 0.8 and 120 M$_{\odot}$. The supernova cut-off mass has been set to 8 M$_{\odot}$ and the black-hole cut-off mass is at 120 M$_{\odot}$. We have considered a metallicity for the models
equal to 0.004, 1/20 times the solar value, which is very close to the metallicity derived for this object, equal to
0.0036. We have considered models with both standard and high mass loss to explore which models fit better the observed properties of the object. 

\subsection{Nebular geometry}

One of the input parameters of the models is the inner radius of the ionized region. In the most successful model, which is the one presented here, the ionization front is located at a distance of 30 pc of the ionizing source. Since the thickness of the H{\sc ii} region reaches approximately 18 pc, the resulting nebular geometry is a thick shell. The total radius, therefore, is not far from that measured on the H$\alpha$ image around the brightest knot, 72$\pm$18 pc, encompassing the same flux measured in the slit and corrected for
aperture effects by G00. This geometry, along with a filling factor of 0.077 leads to a value of the
spherical ionization parameter log $U$ = -2.22, coincident with the value -2.3$\pm$0.1 calculated
by PMD03. This model considers a constant density of 190 particles per cm$^3$, which is the value obtained from the quotient of fluxes of the lines of [S{\sc ii}].

\subsection{Dust content}

We have added a certain amount of dust in order to fit correctly the measured electron temperatures. 
The presence of solid grains within the ionized gas have some consequences on the physical conditions of the nebula
that should not be neglected, including the depletion of metals in gaseous form onto grains or the
absorption of energetic radiation. The heating of the dust can affect, hence, the equilibrium between
the cooling and heating of the gas and, thus,  the electron temperature inside the nebula.
The command PGRAINS, included in the last version of Cloudy, implements
the presence of dust grains as described in van Hoof et al. (2001). This command resolves the grain size
distribution into several bins and treats each bin with its corresponding temperature, potential and drift velocity, which
depend on the diameter of each grain. 
We have assumed the default grain properties of Cloudy 96, which has, essentially,  the properties of the interstellar medium and follows a MRN (Mathis, Rumpl \& Norsieck 1997) grains size distribution.

\begin{figure*}
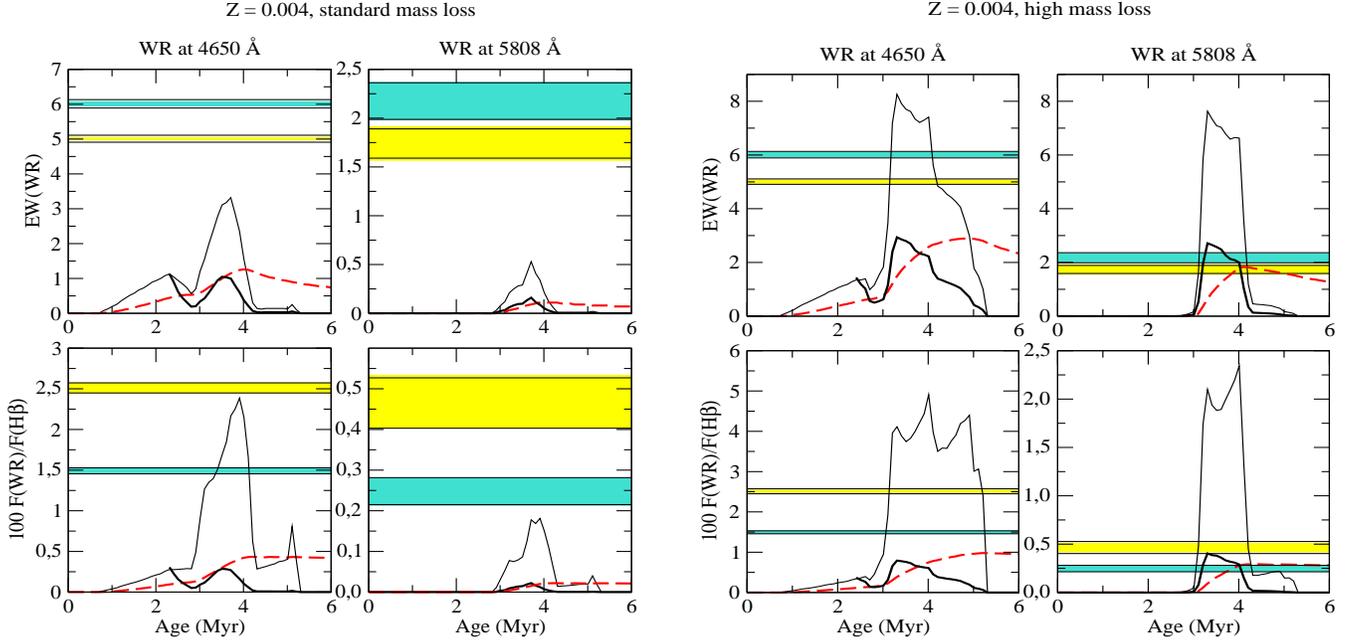

\begin{minipage}{175mm}
\centerline{
\psfig{figure=wrf_0004st_Mrk209.eps,height=8.5cm,width=8.5cm,clip=}
\hspace{0.5cm}
\psfig{figure=wrf_0004hi_Mrk209.eps,height=8.5cm,width=8.5cm,clip=}}
\caption{Comparison between the intensities and equivalent widths of WR red and blue bumps
observed in Mrk209 and those predicted by the models. At left, models with standard mass
loss and at right, with high mass loss. Both have been calculated for a metallicity
Z = 0.004. The clear bands stand for the observed values with their errors as published in G00. 
The dark bands stand for the observed values once corrected for nebular continuum, in the case of
EW and the fraction of absorbed ionizing photons, in the case of relative intensities. 
Models for an instantaneous burst
are plotted as a solid thin black line. Models for a continous star formation are plotted
with a dashed line. Finally, the solid thick line represents models of a composite
population with an instantaneous burst with an age of 1 Myr, resulting in the expected EW(H$\beta$).}

\label{wr0004}
\end{minipage}
\end{figure*}

\begin{figure*}
\begin{minipage}{175mm}
\centerline{
\psfig{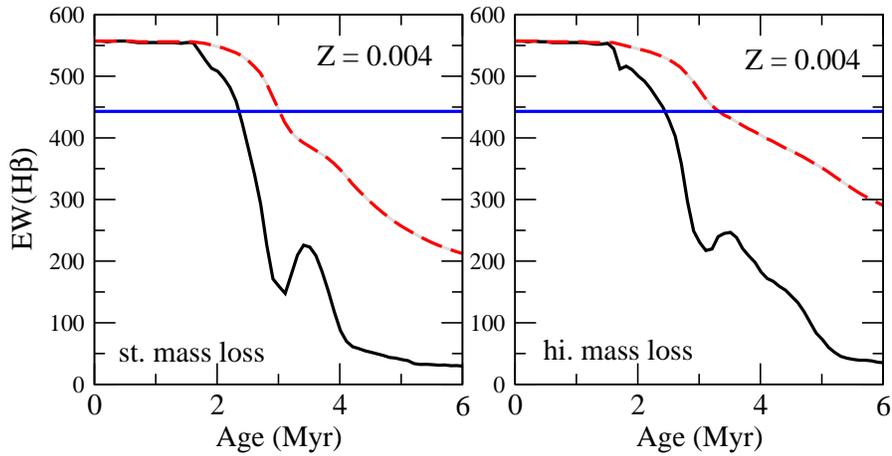}}
\caption{Prediction of the models for the equivalent width of H$\beta$ for both standard and high mass loss and for instantaneous (thick solid line) and continuous (dashed line) star formation rate. The horizontal line represents the observed EW(H$\beta$, once substracted the nebular continuum and corrected for the amount of absorbed ionizing photons as predicted by our model.}
\label{ew0004}
\end{minipage}
\end{figure*}
\begin{figure*}
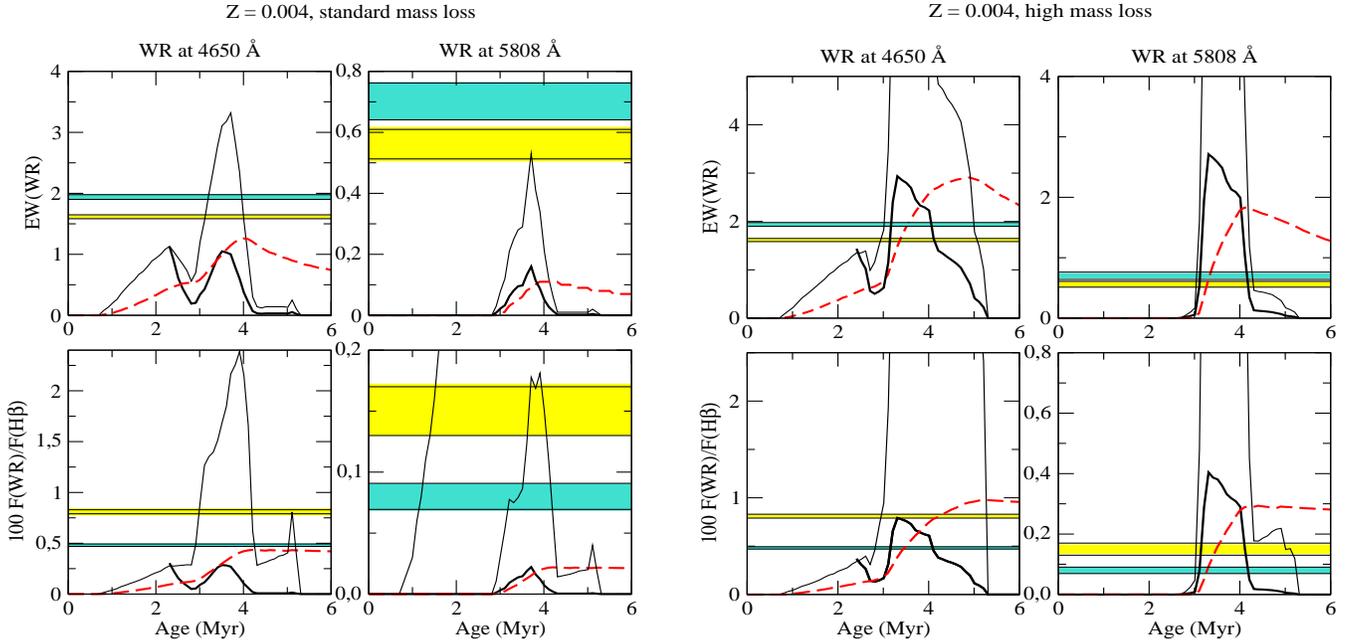

\begin{minipage}{175mm}
\centerline{
\psfig{figure=wrf_0004st_Mrk209b.eps,height=8.5cm,width=8.5cm,clip=}
\hspace{0.5cm}
\psfig{figure=wrf_0004hi_Mrk209b.eps,height=8.5cm,width=8.5cm,clip=}}
\caption{Comparison between the intensities and equivalent widths of the WR red and blue bumps observed and aperture corrected in Mrk209 and those predicted by the models. At left, models with standard mass loss and at right, with high mass loss. Both models have been calculated for a metallicity Z=0.004. The bands stand for the same quantities represented in Figure \ref{wr0004}, but now taking into account the aperture correction. Models for an instantaneous burst are plotted as a solid thin black line. The solid thick line represent models of a composite population with an instantaneous burst with an age of 1 Myr, resulting in the expected EW(H$\beta$) in the slit (443\AA) and that assuming the aperture correction only for the 1.0 Myr population (514\AA)}.

\label{wr0004b}
\end{minipage}
\end{figure*}

\section{Results and discussion}

We find that the model that better fits the observed line intensities, the derived ionic abundances and the other observational properties of this knot is well represented by an instantaneous burst of 1.0 Myr. The inclusion of another burst in the WR phase in order to explain the observed features does not affect the ionization structure of the nebula and the obtained results. This result is equivalent assuming a constant star formation rate for the ionizing burst in such a way that its present age was that in the WR phase.The  predicted emission line intensities relative to H$\beta$ = 100 are presented in column 3 of Table \ref{lines} and can be compared with the observed ones listed in column 2, and with those predicted by the best-fitting model in PMD03. 

We can see all the relevant emission lines are better fitted by the present model than by that presented in PMD03 in which the ionizing radiation was represented by the CoStar atmosphere of a single massive star and a plane-parallel geometry was asumed. Using that model, it was not possible to reproduce the observed T([O{\sc iii}]) due to  the underestimate of the auroral line at 4363 {\AA}. On the other hand, an incorrect simultaneous fitting of the strong lines of [O{\sc ii}], [O{\sc iii}], [S{\sc ii}] and [S{\sc iii}] caused an overestimate of the $\eta'$ parameter, defined as
 \[ \eta' = \frac{[O{\sc ii}]3727/[O{\sc iii}](4959+5007)}{([S{\sc ii}](6717+6731)/[S{\sc iii}](9069+9532)} \]
\noindent (V\'{\i}lchez \& Pagel, 1988), which is indicative of the temperature of the radiation field. 
In the present model, by means of including a more reallistic ionizing spectral energy distribution, derived from 
the evolutionary synthesis of the ionizing cluster, dust physics and an appropriate geometry, we have been able to reproduce adequately the main emission line intensities. Nevertheless, the model is not able to reproduce the mid-infrared lines observed with ISO, obtaining values for both [S{\sc iii}] 18.7$\mu$ and [S{\sc iv}]  lower than observed by  60\% and 40\% respectively. This illustrates the problem that photoionization models have
in general to correctly reproduce the intensities of sulphur emission lines in this spectral range. In Figure \ref{midir} we represent the quotients of the lines of  [S{\sc iv}] 10.51 $\mu$ / [S{\sc iii}] (9069+9532 \AA), at left and of
[S{\sc iii}] 18.67 $\mu$ / [S{\sc ii}] (6717+6731 \AA), at right, versus the quotient of [O{\sc ii}] 3727 \AA / [O{\sc iii}] (4959+5007 \AA), for a sample of BCDs observed both in the mid-IR with Spitzer (IZw18, UM461 and IIZw40; Wu et al., 2006) and ISO (Mrk209; Nollenberg et al., 2002) and the near IR: UM461, IIZw40 and Mrk209 (PMD03), IZw18 (Garnett, 1992). The comparison with the model presented here and a sequence of CLOUDY models simulating a very young cluster ionizing nebulae of different physical conditions show serious discrepancies with observations for the prediction of the fluxes of the sulphur lines at mid-IR.

\subsection{Inner extinction}

The inclusion of grain physics helps to reproduce the intensity of the 
[O{\sc iii}] at 4363 {\AA}, which depends exponentially on electron temperature. This shows the relevance of the presence of dust in the nebula. The required amount of dust implies a dust-to-mass ratio equal to 6.13$\cdot$10$^{-3}$, almost identical to that found for  the Milky Way (Spitzer 1978). This amount of dust for the column density predicted by the model,
implies a visual extinction of A$_V$ = 0.39 mag and would correspond to a logarithmic extinction at H$\beta$ of 0.19. 
This is higher than derived from the Balmer decrement (C(H$\beta$ = 0.00; ITL97, C(H$\beta$ = 0.06; G00), but consistent with what is expected when comparing the computed ionizing star cluster colour index (B-R) with the observed one.  Assuming that the presence of any older stellar population has not a relevant weight for this emitting knot, we can calculate the effect of the extinction by the dust present in the nebula on the synthesized (B-R) colour. The predicted (B-R) colour index for the synthesized clusters as function of age is shown in Figure  \ref{br0004}. 
The average colour predicted by instantaneous bursts of star formation for  the ages between 2 and 5 Myr  assumed in our models is around -0.05. This is bluer than the colour index measured for the knot which is B-R = 0.28$\pm$0.18. However, according to our model, the predicted colour should be reddened by 0.24 magnitudes for the required amount of dust, which would make it consistent with observations. 

The effect of dust has to be considered when calculating the number of ionizing photons reaching the gas. In fact, only a fraction $f$ of the number of ionizing photons emitted by the cluster will ionize the gas
\[Q(H) = f\cdot Q_0(H)\]
\noindent  For the amount of dust required to reproduce the emission line intensities in our model, we obtain $f\approx0.58$. 

\subsection{ionizing stellar populations}

The comparison between the observed intensities and equivalent widths of the blue and red bumps
and those predicted by the models is shown in Figure \ref{wr0004}. The light-coloured  bands represent the 
intensities and the equivalent widths of the WR features with their corresponding errors as 
measured by G00 inside the slit. The dark-coloured bands represent the same quantities corrected for different effects: 
In the case of equivalent widths, for the contribution of the nebular continuum that is 20\% of the ionizing continuum at the wavelenght of the blue bump, and 25\% at that of the red bump as predicted by our model. For the relative fluxes, the correction is due to the absorption of ionizing photons by dust, affecting the intensity of H$\beta$. Regarding models the solid thin line represents 
the predictions by a model with a unique instantaneous burst whose age is 
that shown by the X axes, and the dashed line the prediction for a burst with continuous star formation. 
Agreement between the observed and predicted relative intensities and equivalent widths of the WR features is seen only for the models with a high mass loss rate and for an instantaneous burst of aproximately 3.0 Myr. 

Another important constraint to the model is the equivalent width of H$\beta$, which is an estimator of the
age of the ionizing burst. We consider as well negligible the contribution of the underlying stellar population in this knot with a very high value of EW(H$\beta$).  Again the value of 206 {\AA} reported by G00 has to be corrected for the contribution
of nebular continuum and absorption of ionizing photons by dust. These effects increase the value of  EW(H$\beta$) 
produced by the cluster to 443 {\AA}. This value is much higher than predicted by Starburst 99 models for a cluster of 3.0 Myr and high mass loss rate as it can be seen in Figure \ref{ew0004}. However, this disagreement would be solved  if a younger population is present.
We have therefore explored this possibility by assuming that the ionizing population is composed by two different bursts, one responsible for the WR emission and another that we have taken of 1.0 Myr that would provide the necessary excess of ionizing photons. The predicted values of WR equivalent widths and fluxes for this composite population whose EW(H$\beta$) is that observed and corrected, are shown by solid thick lines in Figure \ref{wr0004}.
It can be seen that no agreement is now found for the blue bump although the red bump values are well reproduced. 

Part of this disagreement  could be due to aperture effects. G00 point to a factor of 3.1 between the H$\beta$ flux measured inside the slit and the total H$\beta$ emission. If we assume that correction and consider the WR stars to be concentrated inside the slit, we obtain the fluxes and EW of the WR features shown in Figure \ref{wr0004b}. 
The measured fluxes and equivalent widths have been divided by the aperture correction factor, as a consequence of the increment of the emission of H$\beta$ and its continuum and assumming it affects similarly to both quantities. Again, we plot the same quantities corrected for the presence of the nebular continuum and the absorption of ionizing photons by dust. The predictions of the models are plotted with a solid thin line for a unique burst of star formation and a dashed line for continuous star formation.  The solid thick line corresponds to a composite population  with an instantaneous burst of the age showed at the X axis, plus a  population of 1.0 Myr, the two of them together providing the expected EW(H$\beta$). Now, we find agreement for both blue and red bump features in the high mass loss rate case assuming either a constant star formation rate at an age of 3.6 Myr or a populations composed of a young cluster of 1.0 Myr and an older one with age 3.2 or 4.2 Myr which contains the WR stars. In the case of the continuous star formation rate, the agreement occurs for an age slightly higher than that compatible for the expected EW(H$\beta$), but probably within the errors.

We can calculate the mass of ionizing clusters in each case, using the number of ionizing photons, which can be obtained
from the flux of H$\alpha$, from the relation:
\[\log Q(H) = 11.86 + \log L(H\alpha)\]
Taking into account the fraction, f, of photons absorbed by dust, the resulting number of ionizing photons is 10$^{51.99 \pm 0.13}$. Neglecting the contribution to the number of ionizing photons by the underlying population :
\[Q(H) = M_{yc} \cdot Q(H)_{yc} + M_{WR} \cdot Q(H)_{WR}\]
where $M_{yc}$ and $M_{WR}$ are the masses of the clusters of  1.0 Myr and in the WR phase, respectively, and
$Q(H\alpha)_{yc}$ and $Q(H\alpha)_{WR}$, their corresponding numbers of hydrogen ionizing photons per unit 
mass.  
The relative contributions of each of the clusters to the total ionizing mass can be derived from the observed equivalent width of H$\beta$ using the equation below: 

\[\frac{EW(H\beta)_{obs}-EW(H\beta)_{yc}}{EW(H\beta)_{WR}-EW(H\beta)_{obs}}=\frac{L_C(H\beta)_{WR}}
{L_C(H\beta)_{yc}} =\]
\[ = \frac{L_C(H\beta)_{WR}/m}{L_C(H\beta)_{yc}/m}\frac{M_{WR}}{M_{yc}}\]

\noindent
where EW(H$\beta$)$_{obs}$ stands for the observed value of the equivalent width of H$\beta$,  EW(H$\beta$)$_{yc}$,  
EW(H$\beta$)$_{WR}$ are the predicted equivalent widths for each of the clusters and L$_C$(H$\beta$)$_{yc}$/m, L$_C$(H$\beta$)$_{WR}$/m are the model predicted continuum luminosities per unit mass at the H$\beta$ wavelength for each of the clusters.

The characteristics of the model ionizing clusters are listed in Table \ref{clusters}. The luminosities for the single ionizing clusters are  given per unit solar mass. Using these values and the expresions above, it is possible to calculate the cluster masses for each assumed case. For the composite population with clusters 1 and 3.2 Myr old, masses of 8.7 $\times$ 10$^4$  and 2.4 $\times$ 10$^4$ M$_{\odot}$ are found. For the 1 and 4.2 Myr old clusters, the corresponding values are 9.3 $\times$ 10$^4$ and 3.3 $\times$ 10$^4$ M$_{\odot}$. This gives a total mass for the ionizing population of 1.11 $\times$ 10$^5$ and 1.26 $\times$ 10$^5$ M$_{\odot}$ respectively. In both cases the younger cluster makes up more than 70 \% of this total mass and dominates the ionization providing more than 80 \% of the hydrogen ionizing photons and about twice the continuum luminosity at H$\beta$ than the cluster containing most of the WR stars. The observed luminosities of the blue and red bumps are better reproduced by the combination with the younger WR cluster, as well as the equivalent widths of the WR features.
On the other hand, a larger cluster mass value is found for the continuous star formation scenario: 6.0 $\times$ 10$^5$ M$_{\odot}$ with  the masses derived from the WR blue and red luminosities being barely consistent with observations. 


\begin{table*}
\centering
\normalsize
\caption{ Characteristics of the ionizing clusters as derived using the calculated number of ionizing photons and  equivalent width of H$\beta$.}

\begin{tabular}{cccccccc}
\hline
\hline
                 & Young  & \multicolumn{2}{c}{WR}      & \multicolumn{2}{c}{Composite}  & Continuous  &   Observed \\
                 & cluster & \multicolumn{2}{c}{cluster}  & \multicolumn{2}{c}{Population } &    SF           &   (corrected)\\
& (per M$_{\odot}$)   & \multicolumn{2}{c}{(per M$_{\odot}$)} &  \multicolumn{2}{c}{}   & (per M$_{\odot}$) &           \\
Age (Myr)   & 1         &         3.2 & 4.2                  &          1, 3.2  & 1, 4.2               &     3.6          &                  \\
\hline
logL(H$\beta$) $^a$        & 34.62 & 34.47 & 34.16 & 39.64 & 39.61 & 33.85 & 39.63 $\pm$ 0.13 \\
logL(WR, 4650)$^a$        & 31.27 & 33.07 & 32.70 & 37.46 & 37.22 & 31.60 & 37.57 $\pm$ 0.13 \\
logL(WR, 5808)$^a$        & --       & 32.71 & 31.81 & 36.57 & 36.32 & 31.11 & 36.74 $\pm$ 0.16 \\
EW(H$\beta$) (\AA )       & 555   &   220  &  167  &  443   &  443   &   427  &        443 \\
EW(4650) (\AA )             & 0.22  &  2.19  &  4.84 &   2.40  & 1.40  &    2.03 & 1.94 $\pm$ 0.04 \\   
EW(5808) (\AA )             &   --    &  7.63  &  1.19 &   0.72  &  0.13  &   1.23 & 0.70 $\pm$ 0.06 \\

\hline
\hline
\multicolumn{8}{l}{$^a$in units of erg $\cdot$ s$^{-1}$}  \\

\end{tabular}

\label{clusters}
\end{table*}

The found ranges for the mass of the starburst lie in a regime where stochastic fluctuations 
of the number of WR stars in relation to the expected value can still be important. In fact, the dispersion
calculated by Monte Carlo simulations (Cervi\~no et al., 2002) for the total mass of the clusters
reach 0.01 in F(WR)/F(H$\beta$) and 1 {\AA} in the EW(WR) for both the blue and red bumps. This can
affect the expected age of the cluster in the WR phase, but the condition of fitting at same time the equivalent width of H$\beta$ constrains the possible age to only 1 Myr around the obtained value.
Nevertheless, we do not find serious discrepancies between  the prediction of both scenarios, a composite population and a continous star formation about the number of WR stars, which is approximately 4 or 5 WC stars and 1 WN star. and those obtained from comparing the luminosities found for both the blue and the red bump, with the mean luminosities of WC and WN stars.

\subsection{Line temperatures, ionic abundances and ionization correction factors}

\begin{figure*}
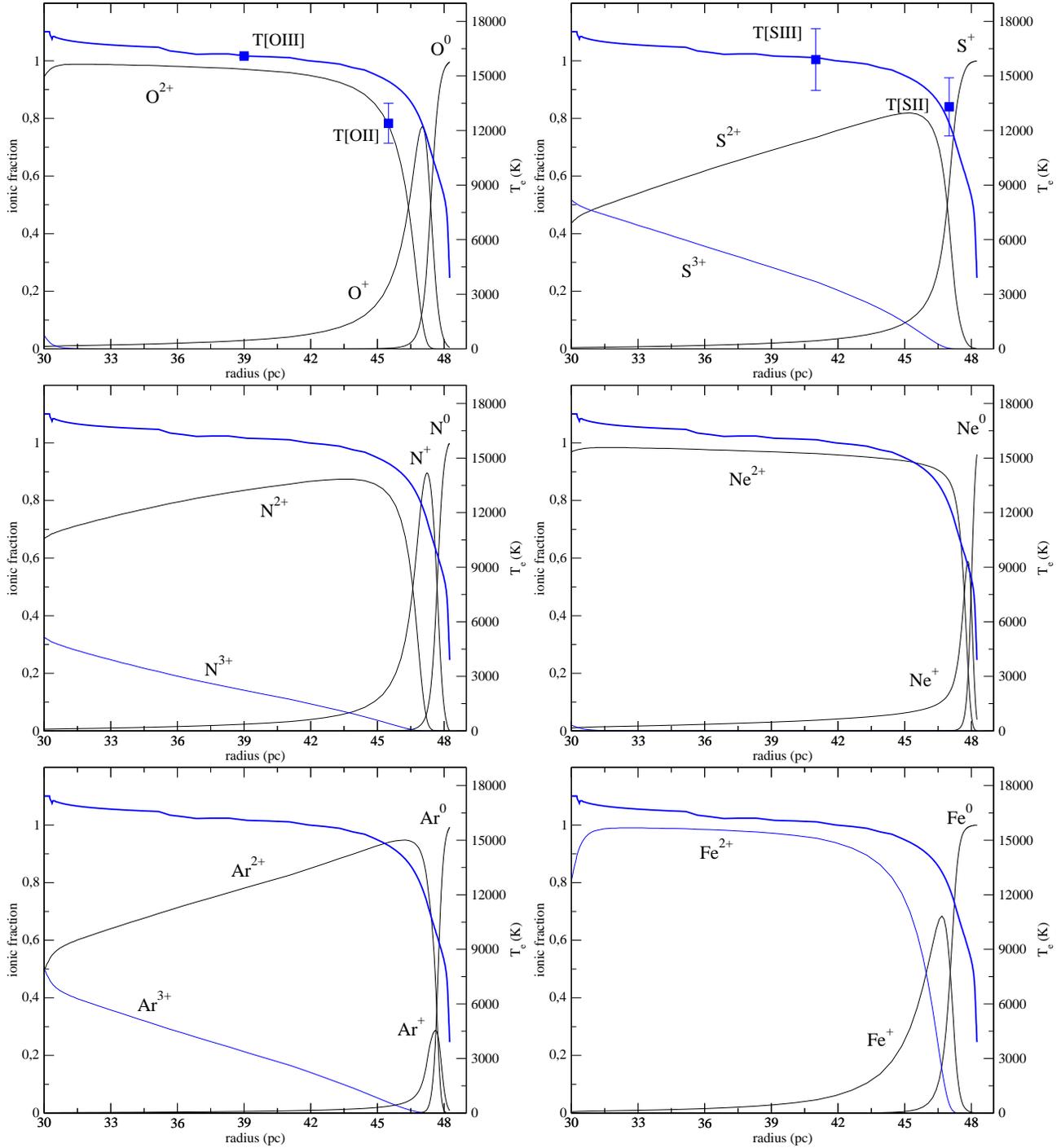

\begin{minipage}{175mm}
\centerline{
\psfig{figure=209_O.eps,width=8.5cm,clip=}
\psfig{figure=209_S.eps,width=8.5cm,clip=}}
\centerline{
\psfig{figure=209_N.eps,width=8.5cm,clip=}
\psfig{figure=209_Ne.eps,width=8.5cm,clip=}}
\centerline{
\psfig{figure=209_Ar.eps,width=8.5cm,clip=}
\psfig{figure=209_Fe.eps,width=8.5cm,clip=}}
\caption{Radial profiles of the relative abundances of the most representative ionic species: oxygen, sulphur, nitrogen, neon, argon and iron. The radius indicates the distance to the ionizing cluster, with the ionization front located at 30 pc.The ionic fractions are marked on the left {\em y} axes. The radial profile of the electron temperature is plotted for all the species along with their available measured values, which have been located at a distance that is a weighted average of the ionic fraction of the corresponding element. Its scale is marked on the right {\em Y} axes in all the plots.}
\label{grad}
\end{minipage}
\end{figure*}

The best-fitted model for the emission lines have been obtained by varying the properties of the
geometry and the metal content of the gas for the obtained stellar energy distribution. Here we
summarize for each ionic species the values of electron temperature and ionic abundances predicted
by this model. In figure \ref{grad} we
represent the radial profiles of the abundances of each ionic species in the best-fitted model along with the
electron temperatures in order to illustrate the most representative temperature in the zones 
where each ionic species stays.

In order to calculate averages of the electron temperature, ionic fractions and ionization correction
factors, we have considered volumetric integrations which fit better the observed values for models with a spherical geometry (Luridiana et al., 2002)

\[\left(\frac{X^i}{X}\right) \equiv \int_0^{R_S}N(X^i)N_edR \Bigg/  \int_0^{R_S}N(Xi)N_edR  \]

The values obtained by the best-fitted model, along with the values obtained from the observations
are tabulated in Table \ref{ab_mod}.\\

\begin{table}
\begin{minipage}{85mm}
\vspace{-0.3cm}
\normalsize
\caption{Observed and model-predicted  electron temperatures and ionic abundances.}

\begin{center}
\begin{tabular}{ccc}
\hline
\hline
  &  Measured & Model-predicted   \\
\hline
T([O{\sc ii}]) (K)                      & 12400$\pm$1100 & 13930 \\
12+log($O^+/H^+$)               &  7.10$\pm$0.11 & 6.88  \\
T([O{\sc iii}]) (K)                     & 16200$\pm$100 & 15850 \\
12+log($O^{2+}/H^+$)            &  7.68$\pm$0.01 & 7.74  \\
12+log(O/H)   &    --          & 7.80  \\
\hline
T([N{\sc ii}])  (K)                     & --             & 13520 \\
12+log($N^+/H^+$)               &  5.64$\pm$0.11 & 5.50  \\
ICF(N$^+$)                          &                          &   9.38    \\
12+log(N/H)                     &   --           & 6.47  \\
\hline
T([Ne{\sc iii}])  (K)                     & --             & 15600 \\
12+log($Ne^{2+}/H^+$)               &  7.01$\pm$0.01 & 6.98  \\
ICF(Ne$^{2+}$)                          &                          &   1.05    \\
12+log(Ne/H)                     &   --           & 7.00  \\
\hline
T([S{\sc ii}]) (K)                      & 13300$\pm$1600 & 13270 \\
12+log($S^+/H^+$)               &  5.12$\pm$0.11 & 5.05  \\
T([S{\sc iii}]) (K)                     & 15900$\pm$1700 & 15600 \\
12+log($S^{2+}/H^+$)            &  5.86$\pm$0.10 & 5.94  \\
12+log($S^{3+}/H^+$)            &  5.93$\pm$0.20 & 5.52  \\
12+log($S^{3+}/H^+$)\footnote{Assuming a constant S$^{3+}$/S$^{2+}$ ratio. See text for details.}            &  5.68$\pm$0.23 & 5.52  \\
ICF(S$^+$+S$^{2+}$)                          &                          &  1.31      \\
12+log(S/H)                     &                & 6.11  \\
\hline
T([Ar{\sc iii}]) (K)                      &             & 15350 \\
12+log($Ar^{2+}/H^+$)            &  5.31$\pm$0.09 & 5.32  \\
ICF(Ar$^{2+}$)                          &                          &   1.22    \\
T([Ar{\sc iv}]) (K)                     &               & 15995 \\
12+log($Ar^{3+}/H^+$)            &  4.49$\pm$0.05 & 4.59  \\
ICF(Ar$^{2+}$+Ar$^{3+}$)                          &                          &    1.02   \\
12+log(Ar/H)                     &                & 5.40  \\
\hline
T([Fe{\sc iii}]) (K)                     &               & 14450 \\
12+log($Fe^{2+}/H^+$)            &  4.92$\pm$0.22 & 5.01  \\
ICF(Fe$^{2+}$)                          &                          &   7.59    \\
12+log(Fe/H)                     &                & 5.89  \\
\hline
\hline
\end{tabular}

\end{center}

\label{ab_mod}
\end{minipage}
\end{table}

\noindent {\bf Oxygen.} Both [O{\sc ii}] and [O{\sc iii}] electron temperatures are relatively well reproduced by the model. While T([O{\sc iii}]) results 2\% lower than the measured value, T([O{\sc ii}]) is
overestimated by about 10\%. However, the lines involved in the determination of this temperature are widely separated  and the measured error, mostly due to uncertainties in the reddening correction, is large. Using the measured [O{\sc iii}] temperature and the grid of models from PMD03, assuming the electron density obtained from [S{\sc ii}] lines, we would obtain a value for T([O{\sc ii}]) equal to 14300 K leading to a value for 12+log(O$^+$/H$^+$) equal to 6.89$\pm$0.11, which is more coincident with the result predicted by the model. 
Although the amount of O$^+$ is not especially relevant for the total abundance of oxygen for this object,
due to its high degree of excitation, this amount becomes more important
for the calculation of the ionization correction factors of some other species through the O$^+$/O 
quotient.  The model gives as well a value of T([O{\sc iii}]) via the recombination lines allowing
to estimate the fluctuations of temperature inside the nebula, which could occasionate an underestimate
of the chemical abundance (Peimbert, 1967). The value of 16100 K for this temperature leads to a $t^2$ = 0.004, which is consistent with the measurements for  Blue Compact Dwarf Galaxies made using the temperature
of the Balmer jump (Guseva et al., 2006: H\"agele et al., 2006), which point to values of $t^2$ negligible in these kind of objects.

The total oxygen abundance is calculated under the assumption that the fractions of neutral hydrogen and oxygen are equal:
\[\frac{O^0}{O} = \frac{H^0}{H}\]
\noindent due to the charge-exchange reaction $O^++H^0 \rightarrow O^0+H^+$ (see Osterbrock, 1989), and this
allows to calculate O/H as:
\[\frac{O}{H} = \frac{O^++O^{2+}}{H^+}\]
In our model  the neutral fraction of both elements are almost coincident: 0.021 for H$^0$/H and 0.019 for O$^0$/O. \\

\noindent {\bf Nitrogen}. T([N{\sc ii}]) can be determined directly when the auroral line, at 5755 {\AA} is measured. Nevertheless in the most objects, as in the case of Mrk209, this line is not detected with enough signal-to-noise ratio. Therefore, the ionic abundance of N$^+$ is calculated under the assumption that T([N{\sc ii}]) $\approx$ T([O{\sc ii}]), since both ions lie in the low excitaton zone of the nebula. In our model, the electron temperature associated
to the zone of N$^+$ is slightly lower than that of O$^+$ and, in fact, is closer to the value of T([S{\sc ii}]).
The N$^+$/H$^+$ ionic abundance calculated with T([O{\sc ii}]) as derived from the measured  T([O{\sc iii}]) using the grid by PMD03 is 12+log(N$^+$/H$^+$) = 5.51$\pm$0.02, almost identical to the value predicted by our model.

The total abundance of nitrogen is calculated under the assumption that
\[\frac{N^+}{N} = \frac{O^+}{O}\]
For our model, this approximation yields a value of ICF(N$^+$) equal to 8.24, confirming the results predicted by the models in PMD03.  The predicted value is slightly higher than the value obtained for the {\em hot} model from Mathis \& Rosa (1991), which predicts ICF(N$^+$) = 6.44.\\

\noindent {\bf Neon}. The ionic abundance of Ne$^{2+}$ is calculated assuming that T([Ne{\sc iii}]) $\approx$ T([O{\sc iii}]),
what seems well justified in view of the results of our model. 
The total abundance of neon is calculated assuming that
\[\frac{Ne^{2+}}{Ne} = \frac{O^{2+}}{O}\]
\noindent which, for our model yields, respectively 0.95 and 0.85. This difference could be caused by the
growing importance of the charge transfer between O$^{2+}$ and H$^0$. According to Izotov et al. (2004)
this effect is only noticeable when O$^{2+}$/O$^+$ is larger than 8. For this model, this quotient is 7.24, and
the relative fraction of Ne$^{2+}$ is slightly larger. On the other hand, the ICF(Ne$^{2+}$) from Mathis \& Rosa (1987) is equal to 0.87.\\

\noindent {\bf Sulphur}. For this element, there are two observable stages in the optical and the near IR parts of
the spectrum respectively. For S$^+$, it is usual to assume that T([S{\sc ii}]) $\approx$ T([O{\sc ii}]), but in this object it is possible to measure the auroral lines of [S{\sc ii}] at 4069,74 {\AA}. The derived [S{\sc ii}] line temperature is consistent with that of T([O{\sc ii}]) within the errors, which are large, and closer to the value predicted by the model. Thus, the derived and model calculated abundance of S$^+$ are similar. 
Regarding S$^{2+}$, the line temperature agrees rather well with
the value predicted by the model and therefore the predicted ionic abundance is 
within the errors of the measured value. Both derived and predicted [S{\sc iii}] electron temperatures
agree with the value obtained from T([O{\sc iii}]) using the empirical relation derived by H\"agele et al. (2006), that is:
\[
T([S{\sc iii}]) = (1.19 \pm 0.08) T([O{\sc iii}]) - (0.32 \pm 0.10)
\]

The ionization correction factor of sulphur accounts for the fraction of S$^{3+}$ in the inner parts of the nebula. 
The first assumption to calculate it was (Peimbert \& Costero, 1969):
\[\frac{S^++S^{2+}}{S} = \frac{O^+}{O}\]
but several authors have pointed that this aproximation could overstimate the fraction of S$^{+3}$ in high excitation-
nebula. Thus, Barker (1980), proposed a new ICF for S$^+$ and S$^{2+}$:
\[ICF(S^++S^{2+}) =  \left[1-\left(1-\frac{O^+}{O}\right)^\alpha\right]^{-1/\alpha}\]
whose fit for the models by Stasi\'{n}ska (1978) gives a value of $\alpha$ = 3. Later models (Stasi\'{n}ska, 1990) 
have produced a lower value of $\alpha$ = 2.

On the other side mid-IR observations for this object exist (Nollenberg et al. 2002) and allow to measure directly 
the S$^{3+}$ abundance and calculate the ICF(S$^+$+S$^{2+}$). The value found by P\'erez-Montero et al. (2006) is 1.97$\pm$0.62, which corresponds to Barker's formula for a value of $\alpha$ = 1.66. In that same work, a new value of the S$^{3+}$ abundance  is proposed assuming a constant ratio of S$^{3+}$/S$^{2+}$, comparing the S$^{2+}$ abundances in the near-IR (9069.9532 \AA) and the mid-IR (18.71 $\mu$).
For our model, the predicted ICF(S$^+$+S$^{2+}$) is 1.31, slightly higher than the value obtained in PMD03 (1.22) but within the errors of the new calculated value by P\'erez-Montero et al. (2006) which is 1.56$\pm$0.76. The ICF predicted by our model corresponds to a value of $\alpha$ = 2.8, slightly higher than the $\alpha$=2.5 found in P\'erez-Montero et al. (2006). The value obtained frm Mathis \& Rosa (1991) for this ICF is 2.28.\\

\noindent {\bf Argon}. Only the emission lines of [Ar{\sc iii}] at 7137 {\AA} and of [Ar{\sc iv}] at 4713,40 {\AA} are detected, thus allowing to calculate the ionic abundances of Ar$^{2+}$ and Ar$^{3+}$ respectively. The assumptions T([Ar{\sc iii}])
$\approx$ T([S{\sc iii}]) and  T([Ar{\sc iv}]) $\approx$ T([O{\sc iii}]) are well justified in both cases.

Since Ar$^{3+}$ is not usually detected in spectra, the ionization correction factor for argon refers to Ar$^{2+}$. The ICF(Ar$^{2+}$) expression in Izotov \& Thuan (1994) from the models of Stasi\'{n}ska (1990)
yields a value of 2.51, while in our model a value of 1.32 is found. Using the coefficients from Mathis \& Rosa (1991), the ICF is found to be 21.6.  Mart\'\i n-Hern\'andez et al. (2002) have propose the relation:  N$^{2+}$/N$^+$ $\approx$ Ar$^{2+}$/Ar$^{+}$. The result of our model for the first quotient is 7.21, while for the second one it is 41.98, what is fur away from the expected result.\\

\noindent {\bf Iron}. The only visible species in the optical spectrum of this galaxy is Fe$^{2+}$ through the
4658 {\AA} line. For this stage of iron ITL97 propose T([Fe{\sc iii}]) $\approx$ T([O{\sc ii}]). In our model the agreement
between both temperatures is good, although the line temperature of Fe$^{2+}$ is slightly higher.

The ICF proposed by ITL97 from Stasi\'nska's (1990) models is:

\[\frac{Fe}{Fe^{2+}} = 1.25 \cdot \frac{O}{O^+}\]

\noindent what results in a ICF(Fe$^{2+}$) = 10.30, slightly higher than the value predicted by our model, equal to 7.59,  
and the value obtained using the ICF scheme proposed by Rodr\'\i guez \& Rubin (2004):

\[ICF(Fe^{2+})=\left[\frac{O^+}{O^{2+}}\right]^{0.09}\cdot\frac{O}{O^+}\]

\noindent which gives in our model a value of 6.90.

\section{Summary and conclusions}

We present a self-consistent model for the brightest knot of the Blue Compact Dwarf galaxy
Markarian 209 calculated using the photoionization code Cloudy 96. 
In order to fit the more relevant emission lines emitted by the ionized gas, we have
introduced as input parameters some of the observed and deduced properties of this
object, including radius, metallicity, WR features, the H$\alpha$ flux and the relative fluxes of
the more relevant emission lines observed both in the optical and near-IR. 
The addition to the analysis of the far-red emission lines, including [O{\sc ii}]$\lambda$7319,7330 {\AA} and
[S{\sc iii}] 9069, 9532 {\AA}, allows to derive with more confidence the ionic abundances of oxygen and
sulphur, as well as the ionization parameter and the equivalent effective temperature, by means of the  [S{\sc ii}]/[S{\sc iii}] ratio (D\'{\i}az et al., 1991) and the $\eta$ parameter (V\'{\i}lchez \& Pagel,
1988) respectively. 

Our model uses as input the spectral energy distribution produced by newly formed stellar clusters which has been calculated using evolutionary synthesis techniques (Leitherer et al. 1999) and the models atmospheres from Pauldrach et al. (2001) and Hillier \& Miller (1998). A metallicity of of 1/20 of solar (Z=0.004), as corresponding to the O/H abundance derived for this object, has been adopted. The ionization front is located at a distance of 30 pc from the ionizing cluster  and the thickness of the ionized region is approximately 18 pc. The total radius is close to that measured on H$\alpha$ images. The model assumes a constant density of 190 cm$^{-3}$, as obtained from nebular diagnostics. 

The addition of a certain amout of dust was required in order to reproduce correctly the measured electron temperatures. This dust implies a gas/dust ratio of 6.13$\times$10$^{-3}$, identical to that in the Milky Way, and a fraction of absorbed photons of 0.58. The dust is included in the Cloudy model using the command PGRAINS which provides a more reallistic treatment of the physics of the dust in the nebula. The model predicted reddening allows to correct the measured B-R index in the region of the burst to a value in agreement with the predicted colour index for the stellar ionizing populations used in the model. Nevertheless, the logarithmic extinction at H$\beta$ predicted by the model for the included amount of dust, 
C(H$\beta$) = 0.19, is not compatible with the value deduced from the Balmer decrement (C(H$\beta$) = 0.06). 
This could be caused by an irregular distribution of the dust into the gas.

We have used the  flux of H$\alpha$, the equivalent width of H$\beta$ and the fluxes and equivalent widths of the WR features in order to constrain the age and properties of the ionizing cluster.  Agreement between measured and predicted quantities is found for a composite population of two clusters, one very young, with an arbitrary age of 1.0 Myr, and
another one in the WR phase with either 3.2 or 4.2 Myr. A marginal agreement is also found for an extended burst of star formation with 3.6 Myr. This scenario of continuous star formation is compatible with the results found by Terlevich et al. (2004) from the distribution of EW(H$\beta$) in HII galaxies.  The masses deduced for the clusters lay in a range where the effect of stochastic effects in the number of massive stars can cause variations in the predicted quantities but not affecting the main conclusions of our work.

We have compared measured and model predicted electron temperatures. The presence of the dust allows to reproduce the auroral lines of [O{\sc iii}] and [S{\sc iii}] and therefore the electron temperatures for these species, without appealing to any other mechanism of heating in the high-excitation zones of the ionized gas. Regarding the electron temperatures in the low-excitation zone, the temperature of [S{\sc ii}] is well reproduced, while for [O{\sc ii}],  is  slightly overestimated by the model, although it should be bore in mind that its determination carries large observational errors. This could affect the derived O$^+$ ionic abundance although it should not affect the total derived abundance of oxygen, since most of it appears as O$^{2+}$. The electron temperatures for the rest of the ionic species follow the usual assumptions about the inner structure of the nebula and the determination of their chemical abundances are predicted to be within the errors of the measured quantities in most cases. The model predicts no fluctuations of temperature in agreement with results found in Blue Compact Galaxies from measurementsof the Balmer jump (Guseva et al., 2006; H\"agele et al., 2006).

Using the ionization correction factors predicted by the models for the various species, we can calculate their total
chemical abundances and therefore we can refine the derivation of some relevant abundance ratios. For N/O, we obtain -1.33, in agreement with the values expected for a certain amount of nitrogen produced as secondary. For S/O, the model predicts a value of -1.69, consistent with the constant value found for the sample of H{\sc II} galaxies with measurements of the [S{\sc iii}] lines at 9069,9532 {\AA} and which is slightly lower than the value measured for the Sun (P\'erez-Montero et al., 2006). This is also the case for Ne/O and Ar/O for which values of -0.8 and -2.34, respectively, are found.

\section*{Acknowledgements}

We would like to thank M. Castellanos, C. Esteban, E. P\'erez, E. Terlevich and J.M. V\'\i lchez for very interesting dicussions 
and suggestions, R. Garc\'\i a-Benito for his help with the photometric images and the referee, Daniel Kunth, for many valuable suggestions and comments, which have helped us to improve this work. \\
This work has been supported by project AYA-2004-08260-C03-03 of the Spanish National Plan for Astronomy and Astrophysics. Also, partial support from the Comunidad de Madrid under grant S0505/ESP/000237 (ASTROCAM) is acknoledged. AID acknowledges support from the Spanish MEC through a sabbatical grant PR2006-0049 and thanks the hospitality of the Institute of Astronomy of Cambridge where the final version of this paper was written.

\end{document}